\documentclass[conference]{IEEEtran}
\IEEEoverridecommandlockouts

\usepackage{cite}
\usepackage{amsmath,amssymb,amsfonts}
\usepackage{algorithmic}
\usepackage{algorithm}

\usepackage{graphicx}
\graphicspath{{../}}
\usepackage{textcomp}
\usepackage{tikz}
\usepackage{pgfplots}
\usepackage{xcolor}
\usepackage{soul,color}
\usepackage{makecell}
\usepackage[font=small,skip=2pt]{caption}
\usepackage{adjustbox}
\usepackage{hyperref}
\usepackage{array}
\usepackage{paralist}
\usepackage{booktabs}
\usepackage{siunitx}
\usepackage{tablefootnote}
\usepackage{subfigure}
\usepackage{float}
\DeclareSIUnit\ms{ms}
\def\BibTeX{{\rm B\kern-.05em{\sc i\kern-.025em b}\kern-.08em
    T\kern-.1667em\lower.7ex\hbox{E}\kern-.125emX}}

\usepackage[acronyms,nomain,xindy,nowarn]{glossaries}
\makeglossaries
\newacronym{5g}{5G}{Fifth-Generation Wireless Systems}
\newacronym{ai}{AI}{artificial intelligence}
\newacronym{ann}{ANN}{Artificial Neural Network}
\newacronym{ask}{ASK}{amplitude-shift keying}
\newacronym[plural={AWGN}, \glsshortpluralkey={AWGN}]{awgn}{AWGN}{additive white Gaussian noise}
\newacronym{au}{AU}{aerial user}
\newacronym{ap}{AP}{access point}
\newacronym{bec}{BEC}{binary erasure channel}
\newacronym{ber}{BER}{bit error rate}
\newacronym{bler}{BLER}{block error rate}
\newacronym{bpsk}{BPSK}{binary phase-shift keying}
\newacronym{bs}{BS}{base station}
\newacronym{bbu}{BBU}{baseband unit}

\newacronym{cdf}{CDF}{cumulative distribution function}
\newacronym{ccdf}{CCDF}{complementary cumulative distribution function}

\newacronym{cnn}{CNN}{convolutional neural network}
\newacronym{csi}{CSI}{channel state information}
\newacronym{comp}{CoMP}{coordinated multi-point}
\newacronym{cran}{C-RAN}{cloud-radio access network}
\newacronym{cca}{CCA}{cloud coverage area}

\newacronym{dnn}{DNN}{deep neural network}
\newacronym{drl}{DRL}{deep reinforcement learning}
\newacronym{dqn}{DQN}{deep Q-network}
\newacronym{daps}{DAPS}{dual active protocol stack}
\newacronym{dff}{DFF}{Decreasing First Fit}
\newacronym{ecc}{ECC}{error-correcting code}
\newacronym{ecdf}{ECDF}{empirical distribution function}
\newacronym{ee}{EE}{energy efficiency}
\newacronym{ec}{EC}{edge cloud}
\newacronym{cc}{CC}{central cloud}
\newacronym{e2e}{E2E}{end-to-end}

\newacronym{ff}{FF}{feed-forward}
\newacronym{ffnn}{FF-NN}{feed-forward neural network}
\newacronym{fsk}{FSK}{frequency-shift keying}
\newacronym{fs}{FS}{functional split}
\newacronym{fsm}{FSM}{finite state machine}

\newacronym{gm}{GM}{Gaussian mixture}
\newacronym{GOPS}{GOPS}{Giga Operation Per Second}

\newacronym{hmappo}{H-MAPPO}{hierarchical multi-agent proximal policy optimization}
\newacronym{hdrl}{HDRL}{Hierarchical deep reinforcement learning}
\newacronym{hmarl}{HMARL}{Hierarchical Multi-Agent Reinforcement Learning}

\newacronym{iid}{i.i.d.}{independent and identically distributed}
\newacronym{il}{IL}{information leakage}
\newacronym{iot}{IoT}{internet of things}
\newacronym{ilp}{ILP}{Integer Linear Programming}

\newacronym{lb}{LB}{lower bound}
\newacronym{los}{LoS}{line-of-sight}
\newacronym{lstm}{LSTM}{long short-term memory}

\newacronym{mac}{MAC}{multiple access channel}
\newacronym{map}{MAP}{maximum a posteriori}
\newacronym{mc}{MC}{Monte Carlo}
\newacronym{mdp}{MDP}{Markov decision process}
\newacronym{mimo}{MIMO}{multiple-input multiple-output}
\newacronym{miso}{MISO}{multiple-input single-output}
\newacronym{ml}{ML}{machine learning}
\newacronym{mmwave}{mmWave}{millimeter wave}
\newacronym{mop}{MOP}{multi-objective programming problem}
\newacronym{mrc}{MRC}{maximum ratio combining}
\newacronym{msc}{MSC}{modulation and coding scheme}
\newacronym{mse}{MSE}{mean squared error}
\newacronym{msac}{MSAC}{masked soft actor-critic}
\newacronym{mappo}{MAPPO}{multi-agent proximal policy optimization}
\newacronym{madrl}{MADRL}{multi-agent deep reinforment learning}
\newacronym{milp}{MILP}{Mixed Integer Linear Programming}
\newacronym{mec}{MEC}{Multi-access Edge Computing}

\newacronym{nlos}{NLoS}{non-line-of-sight}
\newacronym{nn}{NN}{neural network}
\newacronym{nve}{NVE}{normalized validation error}
\newacronym{ric}{Near-RT RIC}{Near-Real Time RAN Intelligent Controller}
\newacronym{noma}{NOMA}{Non-orthogonal multiple access}
\newacronym{nfv}{NFV}{network function virtualization}
\newacronym{nat}{NAT}{network address translation}
\newacronym{nlp}{NLP}{nonlinear programming}

\newacronym{oru}{O-RU}{O-RAN radio unit}
\newacronym{oran}{O-RAN}{open-radio access network}
\newacronym{odu}{O-DU}{O-RAN distributed unit}

\newacronym{pdf}{PDF}{probability density function}
\newacronym{pwc}{PWC}{polar wiretap code}
\newacronym{ppo}{PPO}{proximal policy optimization}
\newacronym{pf}{PF}{processing function}
\newacronym{qam}{QAM}{quadrature amplitude modulation}
\newacronym{qos}{QoS}{quality of service}

\newacronym{ra}{RA}{rearrangement algorithm}
\newacronym{rbf}{RBF}{radial basis function}
\newacronym{relu}{ReLU}{rectified linear unit}
\newacronym{ris}{RIS}{reconfigurable intelligent surface}
\newacronym{rl}{RL}{reinforcement learning}
\newacronym{rnn}{RNN}{recurrent neural network}
\newacronym{rf}{RF}{random forest}

\newacronym{sac}{SAC}{soft actor-critic}
\newacronym{sgd}{SGD}{stochastic gradient descent}
\newacronym{sgp}{SGP}{secure goodput}
\newacronym{simo}{SIMO}{single-input multiple-output}
\newacronym{sinr}{SINR}{signal-to-interference-plus-noise ratio}
\newacronym{siso}{SISO}{single-input single-output}
\newacronym{snr}{SNR}{signal-to-noise ratio}
\newacronym{svm}{SVM}{support vector machine}
\newacronym{sfc}{SFC}{service function chain}

\newacronym{tvd}{TVD}{total variation distance}
\newacronym{ttr}{TTR}{time to trigger}

\newacronym{uav}{UAV}{unmanned aerial vehicle}
\newacronym{ub}{UB}{upper bound}
\newacronym{urllc}{URLLC}{ultra-reliable low latency communication}

\newacronym{v2v}{V2V}{vehicle-to-vehicle}
\newacronym{v2x}{V2X}{vehicle-to-everything}
\newacronym{vm}{VM}{virtual machine}
\newacronym{vnf}{VNF}{virtual network function}
\newacronym{vne}{VNE}{virtual network embedding}
\newacronym{vr}{VR}{virtual reality}

\newacronym{zoc}{ZOC}{zero-outage capacity}
\setacronymstyle{long-short}
\glsdisablehyper

\usepackage[disable]{todonotes}

\begin{document}

\title{RIPPLE: Lifecycle-aware Embedding of Service Function Chains in Multi-access Edge Computing}

\author{\IEEEauthorblockN{Federico Giarrè, Holger Karl}
\IEEEauthorblockA{\textit{Hasso-Plattner Institute (HPI)}, Digital Engineering Faculty,
University of Potsdam\\
Email: federico.giarre, holger.karl at hpi.de}}

\maketitle

\begin{abstract}
In \gls{mec} networks, services can be deployed on nearby \gls{ec} as \glspl{sfc} to meet strict \gls{qos} requirements. As users move, frequent \gls{sfc} reconfigurations are required, but these are non-trivial: \glspl{sfc} can serve users only when all required \glspl{vnf} are available, and \glspl{vnf} undergo time-consuming lifecycle operations before becoming operational. We show that ignoring lifecycle dynamics oversimplifies deployment, jeopardizes \gls{qos}, and must be avoided in practical \gls{sfc} management. To address this, forecasts of user connectivity can be leveraged to proactively deploy \glspl{vnf} and reconfigure \glspl{sfc}. But forecasts are inherently imperfect, requiring lifecycle and connectivity uncertainty to be jointly considered. We present RIPPLE, a lifecycle-aware \gls{sfc} embedding approach to deploy \glspl{vnf} at the right time and location, reducing service interruptions. We show that RIPPLE closes the gap with solutions that unrealistically assume instantaneous lifecycle, even under realistic lifecycle constraints.
\end{abstract}
\glsresetall
\section{Introduction}
We consider a \gls{mec} network in which users access services with strict latency and availability requirements. Such services are deployed as \glspl{sfc}, i.e., chains of microservices, referred to as \glspl{vnf}. \glspl{vnf} can be instantiated on nearby \glspl{ec} and migrated across them to serve users while meeting stringent \gls{qos} constraints. As \glspl{vnf} migrate, \glspl{sfc} are reconfigured to exploit available instances and comply with users' \gls{qos} requirements. Literature shows that dynamic \gls{sfc} reconfiguration during user mobility significantly reduces service interruptions in strict \gls{qos} requirements scenarios \cite{chen_mobility-aware_2019,vieira_mobility-aware_2024,zhao_mobile-aware_2019,hu_mobility-aware_2022}.

Existing work, however, typically assumes that \glspl{ec} can serve any \gls{vnf} immediately upon request—an unrealistic assumption. In practice, \glspl{vnf} must undergo a sequence of time-consuming operations (e.g., download, deployment, startup) before becoming operational, collectively referred to as the \emph{\gls{vnf} lifecycle} \cite{giarre_surfing_2025,fu_fast_nodate,yu_characterizing_2020}. These delays directly impact embedding and reconfiguration effectiveness. For example, if lifecycle operations are not completed in time for a user handover, service interruptions occur, whereas proactive preparation enables seamless continuity. Lifecycle awareness is particularly critical for \glspl{sfc}, which can serve users only when \emph{all} constituent \glspl{vnf} are simultaneously running and interconnected.

Na\"{\i}vely keeping all \glspl{vnf} active on every \gls{ec} is infeasible due to limited resources. Instead, \glspl{vnf} must be selectively prepared at locations where they are likely to be needed. Even with perfect mobility knowledge -- as often assumed in literature \cite{medeiros_tenet_2024,chen_mobility-aware_2019} -- seamless service cannot be ensured unless lifecycle dynamics are properly managed: knowing where and when a user will connect is insufficient if required \glspl{vnf} are not operational in time. Without lifecycle awareness, \gls{sfc} embedding strategies inevitably incur service disruptions and resource inefficiencies, as deployment decisions are delayed by \gls{vnf} startup times.

This issue remains largely unaddressed in the literature. To fill this gap, we develop a lifecycle-aware \gls{sfc} embedding and reconfiguration approach. Specifically, we make the following contributions:
\begin{itemize}
\item We formulate the \gls{sfc} embedding and reconfiguration problem with explicit consideration of \glspl{vnf} lifecycle dynamics.
\item We propose RIPPLE, a heuristic that leverages user connectivity forecasts to proactively prepare \glspl{vnf} at selected \glspl{ec}.
\end{itemize}

\section{System Model}\label{system}

We consider a system with a set of users $\mathcal{U}$ moving across a substrate network composed of $\mathcal{B}$ \glspl{bs} and $\mathcal{M}$ multiplexing nodes interconnecting \glspl{bs}. With each \gls{bs} and multiplexing node,  an \gls{ec} $e \in \mathcal{E}$ is co-located, providing an array $R_{\text{tot}}$ of available computational resources for \glspl{vnf}.
 We assume that during handovers, users do not experience service interruptions  thanks to paradigms such as \gls{daps} \cite{lee_intelligent_2022}. Each user requests an \gls{sfc} $s_u\in \mathcal{S}$. An \gls{sfc} is a set of \glspl{vnf} $\{v_0...v_n\}$ embedded in the network, i.e. individually deployed and appropriately connected by virtual links, which possibly correspond to multiple hops in the substrate network. We consider \glspl{vnf} to be able to serve multiple users at a time. We assume that the computational resources used by \glspl{vnf} are enough to comply with the provider's service level agreement regardless of user load, impacting neither resource usage nor processing time.

The \gls{e2e} delay between a user and the end of the \gls{sfc} is computed as the sum of per-link delays along the path. Propagation and transmission delays are neglected, as users, \glspl{bs}, and \glspl{ec} are assumed to be in close proximity \cite{firouzi_delay-sensitive_2024}. Only processing and queuing delays are considered. Processing delay is modeled as a constant $t_p$ per node \cite{coll-perales_end--end_2022}.
Queuing delay is modeled as an $M/M/1$ queue for wireless links and an $M/D/1$ queue for wired links \cite{firouzi_delay-sensitive_2024}. Each user $u$ generates traffic at rate $\lambda_u$, and each link $l$ has service rate $\mu_l$ with arrival rate $\lambda_l$ equal to the aggregate traffic of users traversing it. While $\mu_l$ is constant for wired links, it is bounded by Shannon capacity for wireless links \cite{liu_data_2022,firouzi_delay-sensitive_2024}. The total \gls{e2e} delay perceived by user $u$ is then computed over the set of links $\mathcal{L}$ used to traverse the chain. We assume every \gls{vnf} in the requested chain has a constant processing time.

\subsubsection*{Virtual Network Functions Lifecycle}

\begin{table}
    \centering
    \caption{Resources usage at each state}
    \begin{tabular}{lcccc}
        \toprule
        State &  Disk & CPU & Memory  \\
        \midrule
        Descriptor  & x & x & x \\
        Source &  $\checkmark$ & x & x \\
        Image &  $\checkmark$ & x & x \\
        Stopped &  $\checkmark$ & x & x \\
        Paused    & $\checkmark$ & x & $\checkmark$\\
        Running &  $\checkmark$ & $\checkmark$ & $\checkmark$ \\
    
        \bottomrule
    \end{tabular}
    
    \label{tab:usage}
\end{table}

We model the \gls{vnf} lifecycle as a \gls{fsm} which captures the possible states of a \gls{vnf} \cite{giarre_surfing_2025}:
\begin{inparaenum}[(i)]
\item Descriptor: only a descriptor file containing \gls{vnf} metadata is available;
\item Source: source files (e.g., code or modules) are available;
\item Image: a built image is available and ready for deployment;
\item Stopped: the \gls{vnf} is deployed but not running;
\item Running: the \gls{vnf} is running;
\item Paused: the \gls{vnf} is paused.
\end{inparaenum}Transitioning between states takes non-negligible time, depending on the transition. The FSM is sufficiently general to model common virtualized software artifacts, such as containers \cite{stahlbock_optimization_2022} and \glspl{vm}\footnote{\url{https://docs.openstack.org/nova/latest/reference/vm-states.html}}, while remaining extensible. For example, \glspl{vm} are typically distributed as pre-built images, eliminating compilation delays at the cost of longer download times; in this case, the Source state can be skipped.

Resource consumption varies across states, as summarized in \autoref{tab:usage}, and determines how many services can coexist on a given \gls{ec}. We assume that each \gls{ec} always stores a descriptor for every \gls{vnf} without occupying disk space, as descriptors are lightweight text files. Finally, we assume that service-related files (e.g., source code or images) cannot be removed without traversing the FSM.

\section{Problem Formulation}\label{sec:problemFormulation}
To tackle service interruptions during user mobility, we want to minimize the amount of \textit{unsuccessful packets} per user. A packet is unsuccessful if either of the following is true:
\begin{inparaenum}[(i)]
    \item the packet exceeds the \gls{e2e} delay threshold required by the service, or
    \item the packet reaches a \gls{vnf} that is not running, hence it is not processed but dropped.  
\end{inparaenum}
Let $p_u^t$ be the binary variable describing wether the packet sent by user $u$ at time $t$ is unsuccessful ($p_u^t=1$) or not. We can formalize the objective function for this problem as:
\begin{equation} 
    \min \quad \lim_{T\rightarrow\infty} \frac{1}{T}\sum_t^T \sum_u^\mathcal{U} p_u^t
  \end{equation}

Any solution to the problem at any time  is subject to the following constraints on \gls{vnf} deployment and \gls{sfc} embedding:
\begin{inparaenum}[(i)]
    \item Each \gls{vnf} required by users must appear only once in their embedding, and no other \gls{vnf} may be included;
    \item \glspl{vnf} may be part of an embedding only if running;
    \item \glspl{vnf} may be in only one state at a time at each \gls{ec};
    \item \glspl{ec}' resource capacity may not be exceeded;
    \item \gls{vnf} state transitions cannot be interrupted;
    \item Both exceeding the latency limit of the \gls{sfc} and connection to not-running instances of \glspl{vnf} results in an unsuccessful packet.
    
\end{inparaenum}

\section{Proposed Approach}
To minimize packet loss, users must be continuously provided with an \gls{sfc} embedding that meets their \gls{e2e} delay requirements during mobility. However, mobility makes frequent reconfiguration unavoidable. As a result, \glspl{vnf} must be proactively deployed at \glspl{ec} ahead of future handovers.
Yet, running all \glspl{vnf} everywhere is not possible due to the constrained \gls{ec} resources, requiring deployments to target locations where users are likely to connect.
Moreover, \glspl{vnf} cannot be treated independently: changes to a single \gls{vnf} may propagate through the \gls{sfc}, affecting its behavior and availability.
Hence, lifecycle management must be explicitly addressed during embedding, as mismatches between embedding decisions and \gls{vnf} availability can severely impact service continuity.

We decompose the problem into three subproblems:
\begin{inparaenum}[(i)]
\item $\mathcal{P}1$: user connectivity forecasting,
\item $\mathcal{P}2$: lifecycle-aware virtual node embedding, and
\item $\mathcal{P}3$: virtual link embedding.
\end{inparaenum}
Classic instances of $\mathcal{P}2$ and $\mathcal{P}3$ are NP-hard even without lifecycle considerations \cite{fischer_virtual_2013}, motivating heuristic and machine learning–based approaches for near-optimal solutions.
Subproblem $\mathcal{P}1$ is equally critical but fundamentally different: since future user connectivity cannot be known exactly, existing work relies on stochastic models or learning-based forecasts.

RIPPLE’s main contribution is solving $\mathcal{P}2$ while explicitly accounting for \gls{vnf} lifecycle; $\mathcal{P}1$ and $\mathcal{P}3$ are addressed by adapting established solutions from the literature. Connectivity forecasts from $\mathcal{P}1$ guide lifecycle decisions in $\mathcal{P}2$, whose outputs then enable an efficient greedy solution for $\mathcal{P}3$.

By decomposing the problem into $\mathcal{P}1$, $\mathcal{P}2$, and $\mathcal{P}3$, the proposed approach allows individual heuristics to be easily replaced or extended to support additional objectives, such as latency minimization.
\subsubsection*{$\mathcal{P}1$} Existing work often overlooks handover uncertainty and network heterogeneity. Many approaches assume perfect knowledge of future connections or rely on purely location-based associations with the nearest \gls{bs} \cite{ouyang_follow_2018,taleb_follow-me_2019,harutyunyan_latency_2022,medeiros_tenet_2024,zhang_service_2024,vieira_mobility-aware_2024}, which poorly reflect real deployments.

RIPPLE relaxes these assumptions by explicitly modeling uncertainty in user connectivity. The probability that a user connects to a given \gls{bs} during mobility is derived from:
\begin{enumerate}
\item the probability density function $f(L)$ of following a path $L \in \mathcal{L}_u$;
\item the conditional probability $P(b \mid l)$ of connecting to \gls{bs} $b$ at location $l$.
\end{enumerate} As these quantities are not directly observable, they must be estimated. While $f(L)$ cannot be inferred explicitly, user trajectories can be predicted over a \emph{finite} horizon $h$ using a mobility model that extrapolates future positions from the last $k$ observations. The horizon length $h$ is critical: short horizons limit proactive reconfiguration, while longer ones suffer from error accumulation. Since location alone does not uniquely determine handovers, RIPPLE employs a classifier to estimate $P(b \mid l)$ for each candidate \gls{bs}. Together, the mobility predictor and classifier form RIPPLE’s inference chain. Their outputs are used to compute the probability that a user does \emph{not} connect to a given \gls{bs} over horizon $h$ as in \cite{giarre_surfing_2025}.
\subsubsection*{$\mathcal{P}2$} Problem $\mathcal{P}2$ addresses the deployment of \glspl{vnf} onto the substrate network. Due to lifecycle dynamics, such deployments are inherently delayed, requiring policies that explicitly account for lifecycle timings and overheads. Trivial solutions (e.g., co-locating entire \glspl{sfc} on a single \gls{ec}) are typically infeasible due to resource constraints, while individual \gls{vnf} placements must be carefully coordinated, as changes may affect multiple services. We propose a heuristic inspired by the \gls{dff} approach \cite{giarre_exploring_2024,vieira_mobility-aware_2024,askari_latency-aware_2019}, extended to incorporate \gls{vnf} lifecycle dynamics. As in \gls{dff}, candidate substrate nodes are sorted by distance and available resources, and a first-fit policy is applied. RIPPLE deploys \glspl{vnf} starting from the furthest feasible nodes, keeping only the \glspl{sfc} heads mobile.
While placement follows \gls{dff}, the \emph{lifecycle state} of each \gls{vnf} depends on the estimated connectivity probability: higher likelihoods trigger deeper lifecycle progression, while lower ones result in partial or no preparation. To capture multi-tenancy, let $\mathcal{U}v$ be the set of users requiring \gls{vnf} $v$. The probability that $v$ is needed at \gls{bs} $b$ is $P_{v,b}~=~1-\prod_{u\in\mathcal{U}v}~P_{b_u \neq b}$.
These probabilities are aggregated at multiplexing nodes. Let $\mathcal{B}e$ denote the set of \glspl{bs} connected to multiplexing node $e$; the probability that $v$ is useful at $e$ is $P_e = 1-\prod_{b\in\mathcal{B}e}(1-P_{v,b})$. Finally, \glspl{vnf} are deployed at \glspl{ec} in lifecycle states reflecting connection probabilities, starting from the end of the \glspl{sfc} and the furthest feasible \glspl{ec} satisfying latency constraints. To improve fairness and avoid worst cases, mapping proceeds layer-wise along the \glspl{sfc}.
At completion, \glspl{vnf} are placed as far as possible from the \gls{bs} layer, minimizing relocations during handovers. When multiple users are likely to connect to the same \gls{ec}, only the heads of their \glspl{sfc} are prepared at the \gls{bs} layer. All solutions produced by the heuristic satisfy the problem's constraints, though suboptimality may arise from forecast errors or threshold-based lifecycle decisions. 

\subsubsection*{$\mathcal{P}3$} The solution to $\mathcal{P}2$ enables the following assumptions:
\begin{inparaenum}[(i)]
\item \glspl{vnf} are not placed beyond a maximum hop distance from the user;
\item \glspl{vnf} are consolidated on as few \glspl{ec} as possible due to the \gls{dff} strategy.
\end{inparaenum}
Given these assumptions and the absence of additional link-related objectives, a simple greedy heuristic is sufficient \cite{giarre_exploring_2024}. Users are connected to the closest instance of the first \gls{vnf} in their \glspl{sfc}, and each subsequent \gls{vnf} is linked to the closest instance of its successor.
If these assumptions do not hold, alternative greedy heuristics from the literature \cite{zhang_online_2021,zhang_adaptive_2019,medeiros_tenet_2024} can be readily substituted.


\section{Numerical Evaluation}\label{sec:numericalevaluation}

We consider \glspl{vnf} to be present in the network as containers, providing the necessary flexibility during embedding and reconfigurations of \glspl{sfc}. We assume containers to be available to all \glspl{ec} via a common repository. We consider lifecycle transition times as derived from studies by Fu \emph{et al.} \cite{fu_fast_nodate} and Yu \emph{et al.} \cite{yu_characterizing_2020}. In particular, we assume the average download time $\bar{t}_{\text{download}}=12$ s, the time to deploy $t_{\text{deploy}}=100$  ms, the time to start a \glspl{vnf} $t_{\text{start}}=530$ ms, and the time to pause a \glspl{vnf} $t_{\text{pause}}=96$ ms. We consider $t_{\text{build}}$ to be instantaneous as it is considered part of the download process in Docker. We assume $t_{\text{stop}} = t_{\text{start}} $, since no quantitative result was found in the literature. Finally, we consider any delete operation in the \gls{fsm} to be instantaneous. We assume all \glspl{vnf} to require a set of resources $R_v~=~(1,1,1)$, consisting of 1 unit of CPU, 1 unit of memory and 1 unit of disk space in order to run, which remains constant regardless of user load.

\subsubsection*{RIPPLE inference chain}

To estimate probabilities to be used in our approach in a real-world infrastructure, we set up an inference chain. The chain is composed of an \gls{lstm}\cite{herzen_darts_2022} model for mobility forecasting and a  \gls{rf}\cite{pedregosa_scikit-learn_2011} model to determine actual user connection probabilities.
Non-default parameters used to train the forecasting model are learning rate set to 0.001 and lag length, i.e. the amount of past coordinates given to the model, set to 5 \cite{oliveira_forecasting_2021}. Scikit-learn's default configuration was used to train the classifier.

\subsubsection*{Comparison cases,  independent variables, metrics}
We combined the previously described modelling into an event-based simulator.
In this section, we compare RIPPLE's results
with those  of two other approaches:

\begin{itemize} 
    \item \emph{Ideal}: optimally solves a relaxed version of the embedding problems $\mathcal{P}2$ and $\mathcal{P}3$ via a \gls{nlp} solver. This approach sets all transition times to 0, allowing for instantaneous embedding. It is meant to show the best possible results achievable, with ideal assumptions that do not reflect the real world. 
    \item \emph{Reactive}: optimally solves problems $\mathcal{P}2$ and $\mathcal{P}3$ via \gls{nlp} for each timestep. Contrary to Ideal, Reactive does not set transition times to 0, operating under realistic lifecycle assumptions. 
    It is meant to show how highly impactful and crucial it is to  consider lifecycle aspects.
\end{itemize}
Users in these experiments move according to the Gauss-Markov mobility model, requiring one of 4 \glspl{sfc}, composed by 4 \glspl{vnf}, taking each 0.1 ms to process, with an \gls{e2e} delay limit of 1 ms \cite{zhang_online_2021,zhang_adaptive_2019}. All approaches are compared on two different scenarios: 
\begin{inparaenum}[(i)]
\item tree topology and 
\item city topology.
\end{inparaenum} 
We vary the mobility model’s correlation parameter $\alpha$ and the forecasting horizon $h$ as independent variables in our comparisons. Adjusting $h$ allows us to analyze how the forecasting horizon affects the quality of inferred connectivity knowledge.
Varying $\alpha$, both during the inference chain's training and testing, is particularly useful to evaluate RIPPLE’s robustness to different levels of mobility randomness.

Approaches are compared on two metrics: unsuccessful packets and burst length. Unsuccessful packets are computed as defined in \autoref{sec:problemFormulation}, while burst length quantifies the duration of service interruptions caused by \gls{sfc} reconfigurations. Although closely related, burst length offers an additional perspective on service continuity beyond unsuccessful packet.

\subsection{Tree topology}
This topology is formed by 16 \glspl{bs}, connected by 4 multiplexing nodes $\mathcal{M}$, connected to the root of the network. In this topology, 4 users move in the network, and each \gls{ec} has a vector of available resource $R_{tot} =(5,8,10)$.

\begin{figure}
    \centering
    \includegraphics[width=0.8\columnwidth]{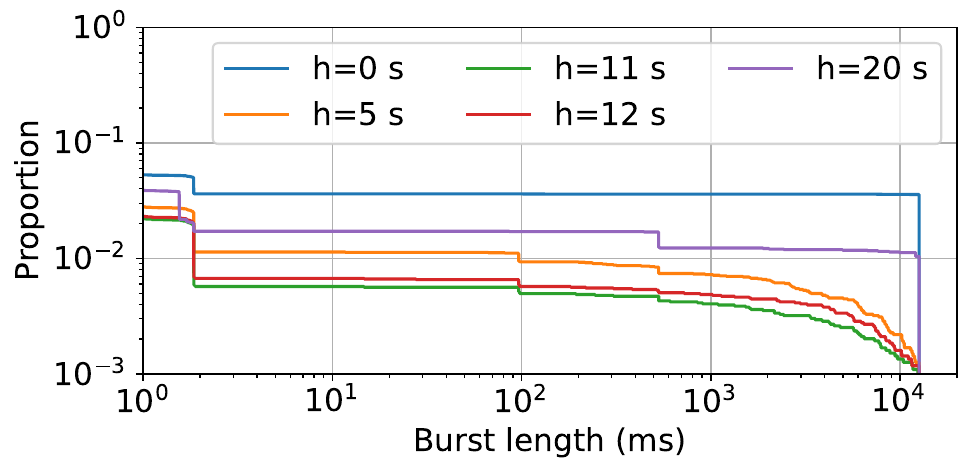}
    \caption{Burst length with respect to an increasing horizon, tree topology. Inference chain trained on $\alpha=0.9$.}
    \label{fig:burst}
  \end{figure}
  


\begin{figure}
    \centering
    \includegraphics[width=0.3\textwidth]{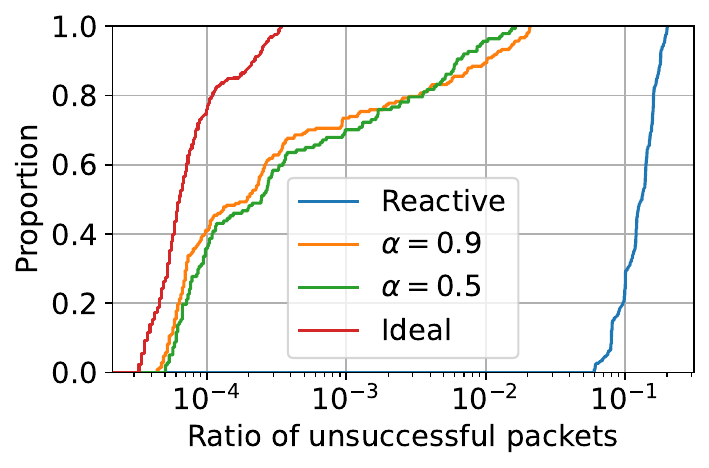}
    \caption{CDF of users with a certain ratio of unsuccessful packets. Inference chain trained on $\alpha=0.9$.}
    \label{ccdf1}
\end{figure}
\begin{figure}
    \centering
    \includegraphics[width=0.23\textwidth]{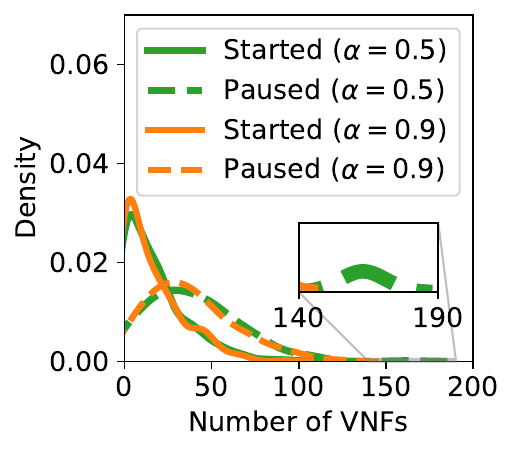}
    \caption{Number of VNFs prepared with respect to different mobility correlation (see legend). Inference chain trained on $\alpha=0.9$.}
    \label{vnfs1}
\end{figure}



\subsubsection*{Burst length} \autoref{fig:burst} shows the \gls{ccdf} of burst lengths for selected horizons. Bursts shorter than 1 ms, caused solely by the wireless channel's conditions, are omitted for clarity. With $h=0$, thus using only the current location estimate, about 3\% of bursts exceed 10 s. As the horizon increases, burst duration decreases, reaching its best at 11 s, where only 0.05\% of bursts last beyond 2 ms. Similar performance (within 0.01\%) occurs when the horizon matches the lifecycle FSM duration. Beyond this point, accuracy declines: for instance, a 20 s horizon yields over 1\% of bursts longer than 1 s.
All subsequent presented results are obtained under the best-performing horizon found.

\subsubsection*{Unsuccessful Packets}
\autoref{ccdf1} presents a \gls{cdf} of the ratio of unsuccessful packets per user. When trained with $\alpha=0.9$, RIPPLE closely matches the Ideal case for over half of the datapoints (within 0.01\%), with a maximum gap below 2\%. To assess robustness, we tested the same model under different mobility patterns: highly correlated ($\alpha=0.9$) and less correlated ($\alpha=0.5$) movement. Performance remains similar across both, though when users move with $\alpha=0.5$, RIPPLE achieves  slightly better results for 20\% of users. As shown in Figure \autoref{vnfs1}, under less correlated movement, RIPPLE starts and pauses \gls{vnf} more frequently, resulting in slightly fewer unsuccessful packets at the cost of higher resources usage.
When trained directly on $\alpha=0.5$, as shown in Appendix \ref{app:1} - \autoref{ccdf2}, RIPPLE further reduces this gap, always at the cost of higher resource usage (see Appendix \ref{app:1} - \autoref{vnfs2}). 

\subsection{City topology}


\subsubsection*{Topology}

We constructed a topology based on the city of Potsdam, Germany, following the method described in \cite{xiang_dataset_2021}. In this experiment, 40 users move through the network. The higher amount of users connecting to \glspl{bs} and multiplexing nodes in the network require for more \glspl{vnf} to be managed at \glspl{ec}. Thus, in this topology the amount of available resources per \gls{ec} is $R_{tot} = (8,10,12)$. This allows \glspl{ec} to proactively prepare \glspl{vnf} that are likely needed without enabling trivial solutions of the problem.

\subsubsection*{Burst Length}
The experiment has been carried out as per the previous topology, without remarkable differences. Results are shown in Appendix \ref{app:2} - \autoref{fig:burstpotsdam}.

\subsubsection*{Unsuccessful Packets}



We tested RIPPLE with the inference chain trained on $\alpha=0.9$. Results are consistent the with previous tests: Appendix \ref{app:2} - \autoref{fig:ccdfPotsdam} shows how RIPPLE closes the gap with the Ideal solver for  more than 50\% of the datapoints. Here, however, users moving with $\alpha=0.5$ have better results than $\alpha=0.9$ for roughly 30\% users. This still comes at the cost of higher VNF replication (see Appendix~\ref{app:2}~-~\autoref{vnfs3}).

\section{Conclusions \& Outlook}
In this paper, we introduced RIPPLE, the first lifecycle-aware \gls{sfc} embedding and reconfiguration approach. By explicitly incorporating lifecycle into the embedding process, RIPPLE substantially reduces the number and duration of service interruptions for users. Furthermore, RIPPLE demonstrated scalability across diverse topologies, ranging from simple tree-like structures to more realistic city-scale networks, while maintaining consistent performance.

These results clearly show that lifecycle awareness is not just an optional enhancement: it is essential for realistic \gls{sfc} embedding. If lifecycle dynamics are ignored, as in the case of existing studies, users are unable to maintain service continuity under realistic assumptions. RIPPLE closes this gap, demonstrating that lifecycle-aware embedding can achieve results close to the ones of literature's traditional, instantaneous embeddings.

\bibliographystyle{IEEEtran}
\bibliography{lib.bib}
\newpage
\begin{appendices}

\section{Tree Topology}\label{app:1}

\begin{figure}[htbp]
    \centering
    \includegraphics[width=0.4\textwidth]{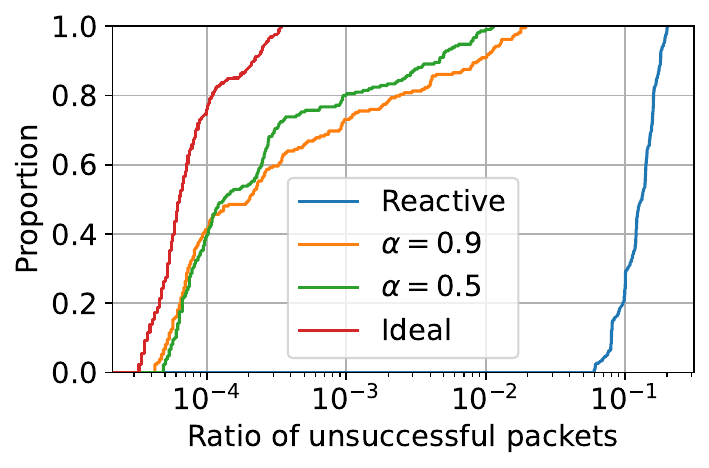}
    \caption{CDF of users with a certain ratio of unsuccessful packets. Inference chain trained with $\alpha=0.5$.}
    \label{ccdf2}
\end{figure}

\begin{figure}[htbp]
    \centering
    \includegraphics[width=0.3\textwidth]{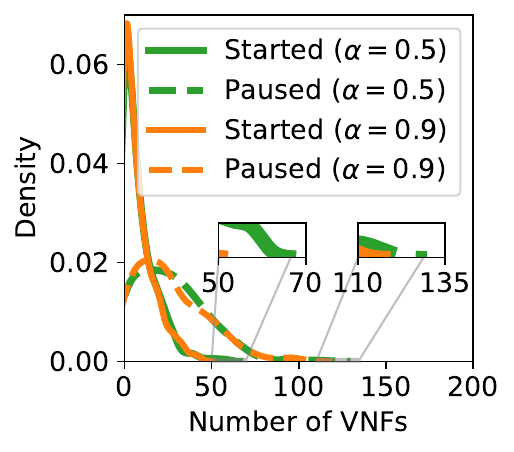}
    \caption{Number of VNFs prepared with respect to different mobility correlation (see legend). Inference chain trained on $\alpha=0.5$.}
    \label{vnfs2}
\end{figure}
\newpage
\section{City Topology}\label{app:2}

\begin{figure}[htbp]
    \centering
    \includegraphics[width=\linewidth]{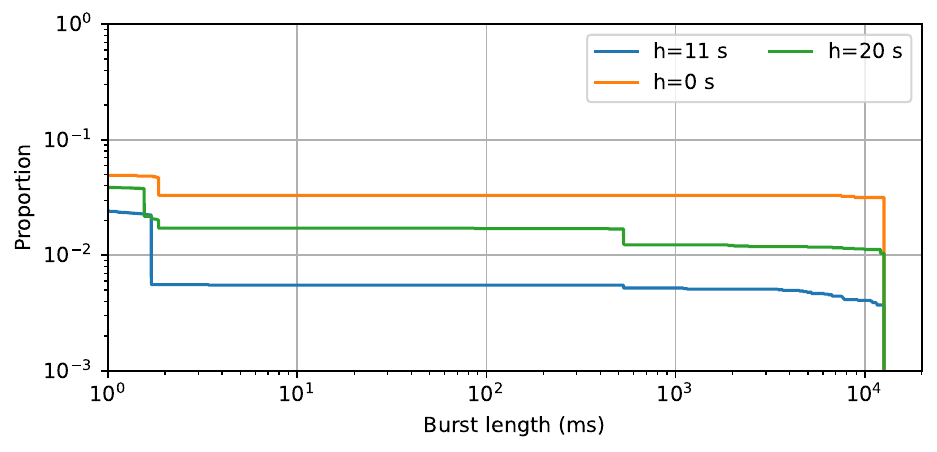}
    \caption{Burst length with respect to an increasing forecasting horizon. Inference chain trained on $\alpha=0.9$}
    \label{fig:burstpotsdam}
\end{figure}
\begin{figure}[htbp]
    \centering
    \includegraphics[width=0.4\textwidth]{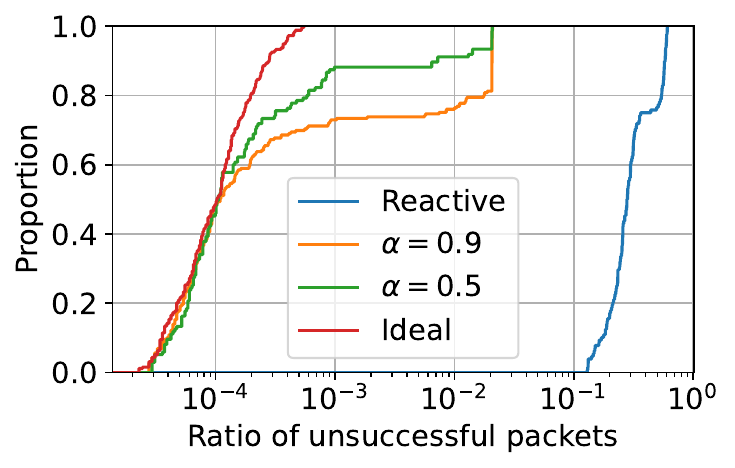}
    \caption{CDF of users with a certain ratio of unsuccessful packets. Inference chain trained with $\alpha=0.9$.}
    \label{fig:ccdfPotsdam}
\end{figure}
\begin{figure}[htbp]
    \centering
    \includegraphics[width=0.3\textwidth]{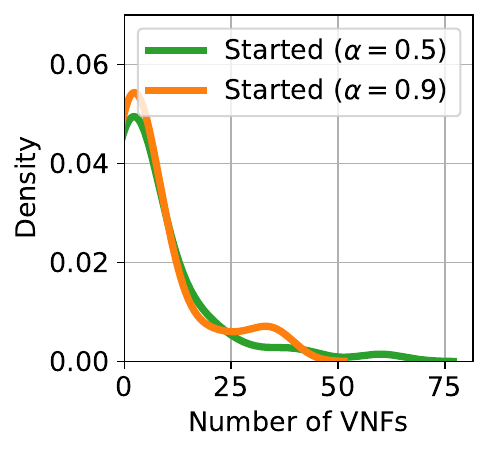}
    \caption{Number of VNFs prepared with respect to different mobility correlation (see legend). Inference chain trained on $\alpha=0.9$. Paused VNFs omitted for readability.}
    \label{vnfs3}
\end{figure}
\end{appendices}

\end{document}